\documentclass[12pt]{article}
\usepackage{graphicx}
\usepackage{amssymb}
\usepackage{amscd}
\usepackage{amsmath}
\usepackage{appendix}
\usepackage{slashed}
\usepackage{epstopdf}

\begin{document}
\begin{titlepage}

{\hbox to\hsize{\hfill June 2017 }}

\bigskip \vspace{3\baselineskip}

\begin{center}
{\bf \large 
Gravitational Instabilities of the Cosmic Neutrino Background with Non-zero Lepton Number}

\bigskip

\bigskip

{\bf Neil D. Barrie and Archil Kobakhidze \\ }

\smallskip

{ \small \it
ARC Centre of Excellence for Particle Physics at the Terascale, \\
School of Physics, The University of Sydney, NSW 2006, Australia \\
E-mails: neil.barrie@sydney.edu.au, archilk@physics.usyd.edu.au 
\\}

\bigskip
 
\bigskip

\bigskip

{\large \bf Abstract}

\end{center}
\noindent
We argue that a cosmic neutrino background that carries non-zero lepton charge develops gravitational instabilities. Fundamentally, these instabilities are related to the mixed gravity-lepton number anomaly. We have explicitly computed the gravitational Chern-Simons term which is generated quantum-mechanically in the effective action in the presence of a lepton number asymmetric neutrino background. The induced Chern-Simons term has a twofold effect: (i) gravitational waves propagating in such a neutrino background exhibit birefringent behaviour leading to an enhancement/suppression of the gravitational wave amplitudes depending on the polarisation, where the magnitude of this effect is related to the size of the lepton asymmetry; (ii) Negative energy graviton modes are induced in the high frequency regime, which leads to very fast vacuum decay producing, e.g., positive energy photons and negative energy gravitons. From the constraint on the present radiation energy density, we obtain an interesting bound on the lepton asymmetry of the universe. 
\end{titlepage}

\vspace{1cm}

\section{Introduction}
Along with the Cosmic Microwave Background radiation (CMB), the existence of the Cosmic Neutrino Background (C$ \nu $B) is an inescapable prediction of the standard hot big bang cosmology (see e.g. \cite{Lesgourgues:1999wu} for a review). It is assumed to be a highly homogeneous and isotropic  distribution of relic neutrinos with the temperature:  
\begin{equation}
T_{\nu}=\left(\frac{4}{11}\right)^{\frac{1}{3}}T_{\gamma}\approx 1.945~{\rm K}~,
\label{1}
\end{equation}
where $ T_{\gamma}=2.725 $ K is the temperature of the CMB today. Unlike the CMB though, the C$ \nu $B is extremely hard to detect and its properties are largely unknown. Namely, the C$ \nu $B may exhibit a neutrino-antineutrino asymmetry
      \begin{equation}
      \eta_{\nu_{\alpha}}=\frac{n_{\nu_{\alpha}}-\bar{n}_{\nu_\alpha}}{n_{\gamma}}=\frac{\pi^2}{12\zeta(3)}\left(\xi_{\alpha}+\frac{\xi_{\alpha}^3}{\pi^2}\right)~,
      \label{2}
      \end{equation}
for each neutrino flavour  $ \alpha=e,~\mu, ~\tau $.  Here $\xi_{\alpha}=\mu_{\alpha}/T$ is the degeneracy parameter, $\mu_{\alpha}$ being the chemical potential for $\alpha$-neutrinos. In fact, such an asymmetry is generically expected to be of the order of the observed baryon-antibaryon asymmetry, $\eta_B=(n_B-\bar n_B)/n_{\gamma}\sim 10^{-10}$, due to the equilibration by sphalerons of lepton and baryon asymmetries in the very early universe. However, there are also models \cite{MarchRussell:1999ig, McDonald:1999in} which predict an asymmetry in the neutrino sector that are many orders of magnitude larger than $\eta_B$. If so, this would have interesting cosmological implications for the QCD phase transition \cite{Schwarz:2009ii} and/or large-scale magnetic fields \cite{Semikoz:2009ye}.     

The most stringent bound on the neutrino asymmetry comes from the successful theory of big bang nucleosynthesis (BBN). BBN primarily constrains the electron neutrino asymmetry. However, this bound applies to all flavours, since neutrino oscillations below $\sim 10$ MeV are sizeable enough to lead to an  approximate flavour equilibrium before BBN, $\mu_e\approx \mu_{\mu}\approx \mu_{\tau} (\equiv \mu_{\nu})$ \cite{Dolgov:2002ab, Wong:2002fa, Abazajian:2002qx}\footnote{See, however, a recent analysis in \cite{Barenboim:2016shh}, where a larger $\eta_{\nu_{\mu}, \nu_{\tau}}$ asymmetry is found to be allowed.}. The updated analysis presented in \cite{Mangano:2011ip} leads to the following bound on the common degeneracy parameter:
\begin{equation}
|\xi_{\nu}|\lesssim 0.049
\label{3}
\end{equation}

In this paper, we argue that the lepton asymmetry in the active neutrino sector leads to gravitational instabilities. These instabilities originate from the gravity-lepton number chiral quantum anomaly, which is present in the Standard Model when considering Majorana neutrinos. Indeed, in the case of Majorana neutrinos, a non-zero lepton asymmetry for active neutrinos implies an imbalance between neutrinos of left-handed chirality and antineutrinos of right-handed chirality which, as we demonstrate explicitly below, leads to the inducement of the gravitational Chern-Simons term in the effective action. This is analogous to the inducement of Chern-Simons terms in gauge theories \cite{Redlich:1984md}.

The induced Chern-Simons term causes birefringence of gravitational waves propagating in the lepton asymmetric neutrino background, which can be sizeable for gravitational waves generated at very early times. More importantly, short-scale gravitational fluctuations exhibit negative energy modes, which lead to a rapid decay of the vacuum state, e.g., into negative energy graviton and photons. Since the graviton energy is not bounded from below, the phase space for this process is formally infinite, that is, the instability is expected to develop very rapidly. Conservatively, we introduce a comoving cut-off $\Lambda$ and compute the spectrum of produced photons as a function of neutrino chemical potential. From the constraint on the radiation energy density today, we then obtain an interesting bound on the neutrino degeneracy parameter:
 \begin{equation}
\xi_{\nu} \lesssim 2\cdot10^{-41}\left(\frac{T_a}{10^{15} \textrm{~GeV}}\right)^{4/3} \left(\frac{M_p}{\Lambda}\right)^{17/3} ~,
\label{4}
\end{equation}  
provided that the lepton asymmetry has been generated above $T_*\gtrsim \frac{440}{\sqrt{\xi}}\sqrt{M_p/\Lambda}$ GeV (here $M_p\approx 2.4\cdot 10^{18}$ GeV is the reduced Planck mass), where $ T_a $ is the temperature at which the asymmetry is generated.

This paper is organised as follows; in Section 2 we describe the calculation of the  graviton polarisation tensor in the presence of a lepton  asymmetric  C$ \nu $B, and consider the associated effective action. Section 3 illustrates the birefringent behaviour of gravitational waves in such a background, while in Section 4 we derive constraints on the C$ \nu $B lepton asymmetry through the induced gravitational instabilities, before concluding in Section 5.

\section{Graviton Polarisation Tensor in the lepton asymmetric C$ \nu $B}
\begin{figure}[t]
		\centering
		\includegraphics[width=0.65\textwidth]{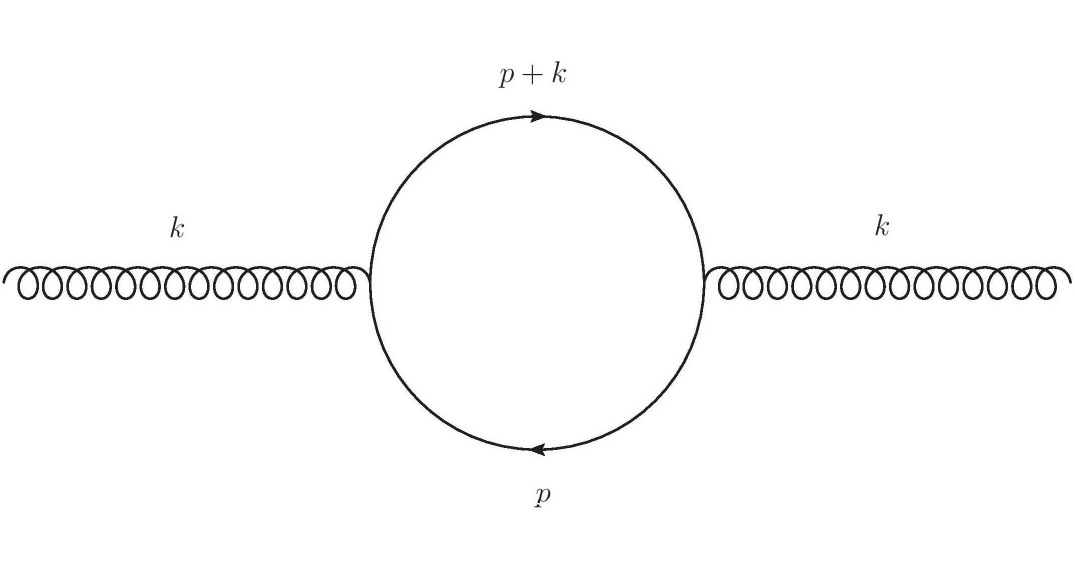}
		\caption{Parity violating contribution to the fermion propagator}
			\label{fig1}
	\end{figure}

We calculate the inducement of the Chern-Simons like terms in the effective graviton Lagrangian through the 1-loop graviton polarization diagram depicted in Figure \ref{fig1}, influenced by a lepton asymmetric neutrino background. The lepton asymmetry is enforced in the Lagrangian through a chiral chemical potential $ \mathcal{L}_{\mu_{\nu}} = \bar{\nu}\slashed{b} \gamma^5 \nu=\mu_{\nu}\bar{\nu}\gamma_0 \gamma^5 \nu$, for which we have considered the frame in which the C$ \nu $B is at rest ($ \slashed{b}=\mu_{\nu}\gamma_0 $). The neutrino propagator is altered as follows:
\begin{equation}
S(p)=\frac{i}{\slashed{p}-m-\slashed{b}\gamma^{5}}=\frac{i}{\slashed{p}-m}\sum_{n=0}^{\infty} \left(-i\slashed{b}\gamma^{5}\frac{i}{\slashed{p}-m}\right)^n\equiv S_0(p)+\sum_{n=1}^{\infty}S_n(p)~,
\label{fermion_prop}
\end{equation}
where $S_0(p)$ is the usual fermion propagator in vacuum. 
The above modified neutrino propagator to first-order in $\mu_{\nu}$ is given by $S(p)\approx S_{0}(p)-i\mu_{\nu}\frac{i}{\slashed{p}-m}\gamma_{0}\gamma^{5}\frac{i}{\slashed{p}-m}$. The higher order terms in $ b_\mu $, or $ \mu $, are neglected because we are only interested in the linear terms in $ b_\mu $, which will result in a Chern-Simons like term. Taking this, and using the standard Feynman rules, we find that the parity odd part of the full graviton polarization tensor is:
\begin{align}
\label{tensor}
\Pi_{\mu\nu\rho\sigma}= & -\int \frac{d^4 p}{(2\pi)^4}(2p+k)_{\nu}(2p+k)_{\sigma}\bigg[Tr(\gamma_{\mu}S_{0}(p+k)\gamma_{\rho}S_{1}(p)) \nonumber \\ & +Tr(\gamma_{\rho}S_{0}(p)\gamma_{\mu}S_{1}(p+k))\bigg]+(\mu\leftrightarrow\nu)+(\rho\leftrightarrow\sigma)+ (\mu\leftrightarrow\nu,\rho\leftrightarrow\sigma)~.
\end{align} 
To evaluate the divergent loop integral in (\ref{tensor}) we employ the dimensional regularization method ($d=4-\epsilon$, $\epsilon\to 0$) and utilise the relations given in Appendix \ref{A}. We hence obtain:
\begin{align}
\Pi_{\mu\nu\rho\sigma}= & \frac{\mu_{\nu}}{8\pi^2}k^{\alpha}\varepsilon_{\mu\rho\alpha 0}\int_{0}^{1} dx \left[\frac{4\pi^2\lambda^2}{M^2}\right]^{\epsilon}\bigg[ 8x^2(1-x)^2(1-2x)^2 \frac{k^2}{M^2}\Gamma(1+\epsilon)k_{\nu}k_{\sigma} \nonumber \\ & +(24x^2-44x+18)\Gamma(\epsilon-1)M^2 \eta_{\nu\sigma}  -16x^2(1-x)^2\Gamma(\epsilon)k^2\eta_{\nu\sigma} \nonumber \\ & -(80x^4 -192x^3 +156x^2 -50x +5)\Gamma(\epsilon)k_{\nu}k_{\sigma} \bigg] \\ & +(\mu\leftrightarrow\nu)+(\rho\leftrightarrow\sigma)+ (\mu\leftrightarrow\nu,\rho\leftrightarrow\sigma)~,
\end{align}
where $ M^2=m^2-x(1-x)k^2 $ and the limit $\epsilon \to 0$ has been assumed. In simplifying this result we find a divergent quantity of the following form:
\begin{equation}
\Pi_{\mu\nu\rho\sigma}^{(div)}= -\frac{1}{\epsilon}\frac{ \mu_{\nu}}{2\pi^2}k^{\alpha}\varepsilon_{\mu\rho\alpha0}  m^2 \eta_{\nu\sigma} +(\mu\leftrightarrow\nu)+(\rho\leftrightarrow\sigma)+ (\mu\leftrightarrow\nu,\rho\leftrightarrow\sigma) ~,
\end{equation}           
where $ \gamma $ is Euler's constant. A straightforward inspection reveals that this divergent term does not satisfy the gravitational Ward identity, $ k^{\nu}\Pi_{\mu\nu\rho\sigma}^{(div)}\neq 0 $, and hence violates the gauge invariance of the effective gravitational action.  This has also been observed previously in a somewhat related calculation in Ref. \cite{Gomes:2008an}. The origin of this violation is rooted in the method of dimensional regularization, which violates Local Lorentz invariance explicitly through the extrapolation to non-integer spacetime dimensions $d=4-\epsilon$. Therefore, following the standard lore, we introduce non-invariant counter-terms to renormalise away this unphysical divergent term.  The polarisation tensor then takes the following simple form:
\begin{equation}
\Pi_{\mu\nu\rho\sigma}= \mu_{\nu} \varepsilon_{\mu\rho\alpha 0}k^{\alpha}[k_{\nu}k_{\sigma}-k^2\eta_{\nu\sigma}]C(k^2) +(\mu\leftrightarrow\nu)+(\rho\leftrightarrow\sigma)+ (\mu\leftrightarrow\nu,\rho\leftrightarrow\sigma)~,
\end{equation}
where 
\begin{equation}
C(k^2)= \frac{1}{192\pi^2} -\frac{m^2}{16\pi^2(k^2)^{3/2}}\left[\sqrt{k^2}-\sqrt{4m^2-k^2}\tan^{-1}\left(\frac{\sqrt{k^2}}{\sqrt{4m^2-k^2}}\right)\right]~.
\end{equation}
This further reduces to:
\begin{equation}
C(k^2) = \begin{cases} -\frac{1}{1920\pi^2} \frac{k^2}{m^2}, & \mbox{if }  k^2/m^2 \ll 1  \\ \frac{1}{192\pi^2}, & \mbox{if } k^2/m^2 \gg 1  \end{cases}
\end{equation}
We wish to investigate the second of these two possible limiting cases, $ k^2/m_{\nu}^2 \gg 1 $. In this limit we obtain the following contribution to the graviton action,
\begin{align}
S_{eff} &= -\frac{\mu_\nu}{192\pi^2 }\int d^4 x\varepsilon_{\mu\rho\alpha 0} h^{\mu\nu}\partial^{\alpha}(\Box h^{\rho\sigma}\eta_{\nu\sigma}- \partial_{\nu}\partial_{\sigma}h^{\rho\sigma}) \nonumber \\
&=  \frac{\mu_\nu}{48\pi^2}\int d^4x~ K^{0}~, 
\label{CSt}
\end{align}
which contains the same number of derivatives as the standard graviton kinetic term in the weak field approximation. In fact, $ K^{0} $ is the linearised 0th component of the four dimensional Chern-Simons topological current:
	\begin{equation}
	K^{\beta}=\varepsilon^{\beta\alpha\mu\nu}(\Gamma^\sigma_{\alpha\rho} \partial_{\mu}\Gamma^\rho_{\nu\sigma}-\frac{2}{3}\Gamma^\sigma_{\alpha\rho}\Gamma^\rho_{\mu\lambda}\Gamma^\lambda_{\nu\sigma}) ~.
	\end{equation}
Therefore, the presence of an asymmetry in the C$ \nu $B replicates Chern-Simons modified gravity:
	\begin{equation}	
	S_{CS}=\int d^4x ~(\partial_{\mu}\theta) K^{\mu} =
	\int d^4x ~ \theta (^*R R)~,
	\end{equation}
where $ \partial_{\mu}\theta=\frac{\mu_{\nu}}{48\pi^2} $.

\section{Gravitational Wave Propagation in asymmetric C$ \nu $B}

The Chern-Simons term in Eq. (\ref{CSt}) is found to induce a birefringence effect on the propagation of gravitational waves. 
The planned gravitational wave detectors, such as eLISA, DECIGO and BBO, can potentially measure the polarization of gravitational waves and hence this birefringence effect.  With this in mind,  we consider the propagation of gravitational waves 
in a lepton asymmetric background over cosmological distances. To this end we parametrise the gravitational waves as: $ h_{ij} = \frac{\mathcal{A}_{ij}}{a(\eta)} \exp[-i(\phi(\eta)-\textbf{k}\cdot \textbf{x})] $, which can be decomposed into the two circularly polarised states: $ e^{R}_{ij}$ and $ e^{L}_{ij}$. 
These two possible circularly polarised states are defined as: $ e^{R}_{ij} =\frac{1}{\sqrt{2}}(e^{+}_{ij}+ie^{\times}_{ij})$ and $ e^{L}_{ij} =\frac{1}{\sqrt{2}}(e^{+}_{ij}-ie^{\times}_{ij})$, which satisfy $ n_{i}\varepsilon^{ijk}e^{R,L}_{kl}=i\lambda_{R,L}(e^{j}_{l})^{R,L} $, where $ \lambda_{R,L}=\pm 1 $. The phase factor $ \lambda_{R,L} $ leads to exponential suppression or enhancement of the left and right circular polarisations of the the propagating gravitational waves, the magnitude of which we shall now calculate. 

From the equations of motion for the action $ S=S_{EH}+ S_{eff} $ \cite{Alexander:2007kv} we obtain:
		\begin{align}
		(i\phi_{,\eta \eta}^{R,L}+(\phi_{,\eta }^{R,L})^2 +\mathcal{H}_{,\eta }& +\mathcal{H}^2-|\textbf{k}|^2)\left(1-\frac{\lambda_{R,L}\kappa\theta_{, \eta} }{a^2}\right)\nonumber \\ &= \frac{i\lambda_{R,L}|\textbf{k}|}{a^2}(\theta_{,\eta \eta}-2\mathcal{H} \theta_{, \eta})(\phi_{,\eta }^{R,L}-i\mathcal{H})~.
		\end{align}
We will first solve the above equation, assuming propagation in the matter dominated epoch $ a(\eta)=a_{0}\eta^2=\frac{a_{0}}{1+z} $. The accumulated phase over the length of propagation, to first order in $ \theta $, is given by,
		\begin{align}
\Delta\phi^{R,L}_{mat}& =i\lambda_{R,L}|\textbf{k}|H_{0}\int_{\eta}^{1}\left[\frac{1}{4}\theta_{,\eta \eta}- \frac{1}{\eta}\theta_{, \eta}\right]\frac{d\eta}{\eta^4} ~.
		\end{align}
In the case considered in this manuscript, we make the following identification $ \theta_{, \eta} =\left(\frac{a(\eta_{0})}{a(\eta)}\right)^2\frac{\mu_0}{48\pi^2 M_{p}^2}$, where $ \mu_0= a(\eta)\mu_{\nu}  $ is the present neutrino chemical potential. For this lepton asymmetric C$ \nu $B,
		  \begin{equation}
		  \Delta\phi_{mat}^{R,L}=-i\frac{1}{288 \pi^2 }\frac{\mu_{\nu} H_{0}}{M_{p}^2}\left(\frac{|\textbf{k}|}{1 ~\textrm{GeV}}\right) (1+z)^4~.
		  \end{equation}
		   Hence the ratio of the amplitudes of each polarisation is given by: 
		 		  \begin{equation}
		 		\frac{h_{R}}{h_{L}}\propto e^{-2|\Delta\phi_{mat}^{R,L}|} ~.
		 		 \end{equation} 
		 	
Taking into account the current bounds on the C$ \nu $B asymmetry parameter, $ \xi $, we find $ |i\Delta\phi^{R,L}|\lesssim 10^{-87} \left(\frac{|\textbf{k}|}{1 ~\textrm{GeV}}\right) $, for $ z\sim 30 $. Therefore, the accumulated phase difference for $ z\sim 30 $ sources is too small to be observable by any conceivable gravitational wave detector. 

The more interesting scenario to consider is the propagation of gravitational waves from sources in the very early in the universe. At early times the chemical potential would have been larger and the longer accumulated propagation time. Conceivably, any early universe sources could provide constraints, if the different polarisations are measurable. Therefore, we now consider gravitational waves produced at very early times, during the radiation dominated epoch after reheating. The accumulated phase now reads:
		\begin{align}
			\Delta\phi^{R,L}_{rad}& =i\lambda_{R,L} \frac{|\textbf{k}|}{\Omega_{\textrm{r,0}}H_0^2} \int_{\eta}^{1}\left[\frac{1}{2}\theta_{,\eta \eta}- \frac{1}{\eta}\theta_{, \eta}\right]\frac{d\eta}{\eta^2} ~,
		\end{align}
		where $ \Omega_{\textrm{r,0}}\sim9.2\cdot10^{-5} $ is the radiation density parameter today. After solving the  integral we find:
		  \begin{equation}
		  \Delta\phi^{R,L}_{rad}\simeq-i\lambda_{R,L}\xi_{\nu}\left(\frac{|\textbf{k}|}{1 ~\textrm{GeV}}\right) \left(\frac{T_s}{1 ~\textrm{TeV}}\right)^4~,
		  \label{rad_rel}
		  \end{equation}
where we have redefined the redshift in terms of the temperature at which the gravitational waves are produced, $ T_s $, or when the asymmetry is generated, whichever is lowest.
		  
From Eq.  (\ref{rad_rel}), it can be seen that if an asymmetry in the C$ \nu $B is present at early times, which equilibrium sphalerons transitions may assure, then it is possible to get significant birefringent behaviour in the propagation of gravitational waves from primordial sources. This is dependent on the momenta $|\textbf{k}|$ of the gravitational waves and size of the asymmetry.

\section{Induced Ghost-like Modes and Vacuum Decay}

Another interesting consequence of the induced Chern-Simons term in Eq. (\ref{CSt})  is that short-scale gravitational fluctuations exhibit negative energy modes, which lead to a rapid decay of a vacuum state e.g., into negative energy graviton and photons \cite{Dyda:2012rj}. Since the graviton energy is not bounded from below, the phase space for this process is formally infinite \cite{Carroll:2003st,Cline:2003gs}, and as such will develop very rapidly. We investigate the production of two photons and a negative energy graviton via this process, to obtain constraints on the neutrino asymmetry at early times. The relevant effective interaction is of the form:
\begin{align}
	S_{\textrm{int}} & \sim  \frac{1}{m_{*}} \int d^4 x h^{can}_{\mu\nu}T^{\mu\nu} \\ & = \frac{1}{m_{*}} \nonumber \int d^4 x \frac{1}{2}h^{can}F_{\mu\nu}F^{\mu\nu}-h^{can}_{\mu\nu}F^{\mu\alpha}F_{\alpha}^{\nu} ~,
\end{align}
where  the canonically normalised graviton field is: $ h^{can}_{\mu\nu}=m_{can}h^{can}_{\mu\nu} $, with 
\begin{equation}
	m_{can}=M_p\sqrt{1+\lambda_{R,L}\frac{|\textbf{k}|}{a m_{CS}}}~,
\end{equation}
where $ m_{CS} $ is the analogous Chern-Simons mass scale:
\begin{equation}
	m_{CS}(t)=\frac{M_p^2}{\mu_\nu}=\frac{M_p^2}{\xi T }=\frac{a(t) M_p^2}{\mu_0} ~.
\end{equation}

\subsection{Photon Energy Spectrum from Induced Vacuum Decay}

To obtain a finite result for the decay rate we need to constrain the phase space. In the absence of a fundamental physical reason for such a truncation, we, following \cite{Carroll:2003st,Cline:2003gs}, simply cut-off the three momenta at $|\textbf{k}|_{max}=\Lambda$.   
 In the analysis that follows, we consider decays into this mode as it will have the largest contribution to the energy density of the generated photons. In addition, we take the reasonable approximation:
\begin{equation}
	m_{can}\simeq \sqrt{\frac{|\textbf{k}|\mu_{\nu}}{a}}~,
\end{equation}
and consider the dynamics of our scenario after reheating and prior to BBN, when the universe is radiation dominated and evolves as follows, 
\begin{equation}
	a(t)=a_0\sqrt{t}= \sqrt{2\Omega_{\textrm{r,0}}^{1/2}H_0 t} ~,
\end{equation}
where $ \Omega_{\textrm{r,0}}\sim9.2\cdot10^{-5} $ is the radiation density parameter today.

The time at which this ghost term is no longer present will be defined as $ t_* $ and can be found in terms of the scale factor:
\begin{equation}
	1\simeq\frac{\Lambda}{a(t_{*}) m_{CS}(t_{*})}~~\Rightarrow~~ a(t_{*})\simeq\sqrt{\frac{\mu_0\Lambda}{M_p^2}} ~~ \textrm{or} ~~ a(t_{*})\simeq\frac{\xi_{\nu} T_{*}\Lambda}{M_p^2}~,
	\label{scale_eq}
\end{equation}
where $ T_{*} $ is the temperature at which the ghost terms stop contributing.

This fixes the time at which the ghost modes no longer exist, and decay of the vacuum ceases. We can reinterpret this as a temperature, so that it is possible to associate this with the maximal reheating temperature, and also ensure it does not  have adverse implications on BBN.
If we assume that the asymmetry is present/produced during the reheating epoch, and prior to BBN, the scale factor has a $ \frac{1}{T} $ dependence, if ignoring the decoupling of radiation degrees of freedom. The scale factor takes the following form:
\begin{equation}
	a(t)\simeq\left(\frac{90\Omega_{\textrm{r,0}}}{g^*\pi^2}\right)^{\frac{1}{4}}\frac{\sqrt{H_0 M_p}}{T}~,
	\label{scale_eq_T}
\end{equation}
where $ g^*=106.75 $. 
Equating Eq. (\ref{scale_eq}) and (\ref{scale_eq_T}) to find the temperature at which this effect ends we find: 
\begin{align}
	T_*&=\left(\sqrt{\frac{90\Omega_{\textrm{r,0}}}{g^* \pi^2}}\frac{H_0 M_p^3}{ \xi_{\nu}^2}\right)^{\frac{1}{4}}\sqrt{\frac{M_p}{\Lambda}} \nonumber \\
&\simeq \frac{440}{\sqrt{\xi_{\nu}}}\textrm{~GeV~}\sqrt{\frac{M_p}{\Lambda}}~.
	\label{T_xi_relation}
\end{align}

Given that the maximum reheating temperature is $ \sim10^{15} $ GeV, 	Eq. (\ref{T_xi_relation}) implies we can constrain the production temperature of neutrino asymmetries to be in the range  $ \xi_{\nu}\gtrsim 2 \cdot 10^{-25}\frac{M_p}{\Lambda}  $, with smaller $ \xi $'s not generating ghost like modes after reheating. We also assume here that $ \xi $ is approximately constant, and hence is the same parameter currently constrained by BBN measurements in the calculation of the lepton asymmetry stored in the C$ \nu $B.

Next we compute the spectrum of photons generated by the induced vacuum decay, and subsequently the energy density, which can be constrained by experiment. It is given by:
\begin{equation}
	\frac{1}{a^3} \frac{d}{dt}(a^3 n(k,t))=\Gamma \delta\left(\frac{|\textbf{k}|}{\Lambda}-1\right)~,
	\label{eq.}
\end{equation}
where $ n(k,t) $ is the number of photons per unit logarithmic wave number $ |\textbf{k}| $ and $ \Gamma $ is the total decay width, which we take to approximately be:
\begin{equation}
\Gamma\sim\frac{\Lambda^6}{m_{can}^2}= \frac{a(t) \Lambda^6}{|\textbf{k}|\mu_{\nu}} = \frac{a(t)^2 \Lambda^5}{\mu_0}~.
\label{rate}
\end{equation}
Since the above decay rate is much faster than the expansion rate of the universe, we may safely assume that the decay happens instantaneously. Therefore, we fix the scale factor in Eq. (\ref{rate}) at time $t_a$, when the asymmetric background is first produced.
We then integrate Eq. (\ref{eq.}) between the time $ t_a $ and when the ghost terms are no longer present $ t_* $:
\begin{equation}
|\textbf{k}|n_{*}(|\textbf{k}|)  \sim \frac{a(t_{*})^2 \Lambda\Gamma_a}{5\Omega_{\textrm{r,0}}^{1/2}H_0}~.
\end{equation}
Taking into account the dilution factor due to the expansion of the universe since the end of photon production to today, $ \left(\frac{a(t_{*})}{a_0}\right)^3=a(t_{*})^3 $, we obtain:
\begin{equation}
|\textbf{k}|n_{0}(|\textbf{k}|)  \sim \frac{a(t_{*})^5 \Lambda\Gamma_a}{5\Omega_{\textrm{r,0}}^{1/2}H_0}~.
\end{equation}
Therefore, the energy density for a given momenta $ k $ is:
\begin{equation}
\frac{dE}{d^3 x d \ln |\textbf{k}|} \sim |\textbf{k}|n_0(|\textbf{k}|)\sim \frac{\xi^4 T_{*}^5 }{10 T^2_a} \sqrt{\frac{M_p^3}{H_0}} \left(\frac{\Lambda}{M_p}\right)^{11}~.
\label{eng_den}
\end{equation}

We can obtain a bound on the energy density of the produced photons, through the observation that the universe is not radiation dominated today:
\begin{equation}
	\frac{dE}{d^3 x d \ln |\textbf{k}|} \lesssim M_p^2 H_0^2~.
	\label{eng_constraint}
\end{equation}
This means we get the following constraint on $ \xi_{\nu} $, assuming the asymmetry is generated above the characteristic temperature $ T_{*} $, when requiring consistency with observation:
\begin{equation}
\xi_{\nu} \lesssim 2\cdot10^{-41}\left(\frac{T_a}{10^{15} \textrm{~GeV}}\right)^{4/3} \left(\frac{M_p}{\Lambda}\right)^{17/3}~,
\label{eng_constraint2}
\end{equation}
for which it is assumed $ T_{a} \gtrsim \frac{440}{\sqrt{\xi_{\nu}}}\textrm{~GeV} \sqrt{\frac{M_p}{\Lambda}} $. Equivalently,
\begin{equation}
T_* \gtrsim 10^{23} \textrm{~GeV} \left(\frac{T_a}{10^{15} \textrm{~GeV}}\right)^{-2/3} \left(\frac{\Lambda}{M_p}\right)^{17/6}~.
\end{equation}
Thus we arrive at the conclusion that, unless $ \Lambda\ll M_p $, the resulting photon energy density from the induced vacuum decay 
can hardly be accommodated with observation.  Substituting the constraint in Eq. (\ref{eng_constraint2}) into that for the asymmetry stored in the C$ \nu $B as a function of $ \xi_{\nu} $, in Eq. (\ref{2}), we find the following bound:
\begin{equation}
	\eta_{\nu}\lesssim 10^{-41}\left(\frac{T_a}{10^{15} \textrm{~GeV}}\right)^{4/3} \left(\frac{M_p}{\Lambda}\right)^{17/3}~.
\end{equation}
If we instead assume that $ T_{a} \lesssim \frac{440}{\sqrt{\xi_{\nu}}}\textrm{~GeV} \sqrt{\frac{M_p}{\Lambda}} $, and hence vacuum decay does not occur, then we get the following constraint on $ \eta_{\nu} $,
\begin{equation}
\eta_{\nu}\lesssim 0.033\left(\frac{ 2000 \textrm{~GeV}}{T_{a}}\right)^2 \frac{M_p}{\Lambda}
\end{equation}
where $ \eta_{\nu}\lesssim 0.033 $ is the current upper limit from BBN constraints.

\section{Conclusions}
In this paper we have argued that the relic neutrino background with non-zero lepton number exhibits gravitational instabilities. Fundamentally, these instabilities are related to the gravity-lepton number mixed quantum anomaly. Indeed, in the relevant limit of vanishing neutrino masses, we have explicitly calculated the parity odd part of the graviton polarization tensor in the presence of a lepton asymmetric C$ \nu $B, which replicates the gravitational Chern-Simons term in the effective action. 

The induced Chern-Simons term leads to birefringent behaviour leading to an enhancement/suppression of the gravitational wave amplitudes depending on the polarisation. While this effect is negligibly small for local sources, we demonstrate that it could be sizeable for gravitational waves produced in very early universe, e.g. during a first-order phase transition.  

In addition to the above, we have also argued that  short-scale gravitational fluctuations in the presence of an asymmetric C$ \nu $B  exhibit negative energy modes, which lead to a rapid decay of a vacuum state e.g., into negative energy graviton and photons. Since the graviton energy is not bounded from below, the phase space for this process is formally infinite, that is the instability is expected to develop very rapidly.  From the constraint on the radiation energy density today, we have obtained an interesting bound on the neutrino degeneracy parameter in Eq. (\ref{4}).  

We believe that the findings reported in this paper will prove to be useful for further understanding the properties of the C$ \nu $B and putting constraints on particle physics models with a lepton asymmetry.    

 \section*{Acknowledgements} We would like to thank Rohana Wijewardhana for useful discussions. This work was supported by the Australian Research Council.

\appendix
\section{Appendix: Further details of calculations}
\label{A}
Useful equations:
\begin{equation}
Tr(\gamma_{\mu}\gamma_{\alpha}\gamma_{\rho}\gamma_{\beta}\gamma^{5})=-4i\varepsilon_{\mu\alpha\rho\beta}
\end{equation}
Taking $ \epsilon \rightarrow 0 $ :
\begin{equation}
	\Gamma(1+\epsilon)|_{\epsilon\rightarrow 0}\simeq 1
,~~~~
	\Gamma(\epsilon)|_{\epsilon\rightarrow 0}\simeq \frac{1}{\epsilon}-\gamma
,~~~~
	\Gamma(\epsilon-1)|_{\epsilon\rightarrow 0}\simeq -\frac{1}{\epsilon}+\gamma-1
\end{equation}

\begin{equation}
	\eta_{\mu\nu}\eta^{\mu\nu}\simeq 4-2\epsilon
\end{equation}

\begin{equation}
	\left[\frac{4\pi\lambda^2}{M^2}\right]^\epsilon |_{\epsilon\rightarrow 0} \simeq 1+\epsilon\ln\left(\frac{4\pi\lambda^2}{M^2}\right)
\end{equation}
\\

Dimensional regularisation of loop integrals:

\begin{equation}
i\int \frac{d^N p}{(2\pi)^N}\frac{1}{(p^2-m^2)^2} = -\frac{1}{16\pi^2}\left[\frac{4\pi^2\lambda^2}{M^2}\right]^{\epsilon}\Gamma(\epsilon)
\end{equation}

\begin{equation}
i\int \frac{d^N p}{(2\pi)^N}\frac{1}{(p^2-m^2)^3} = \frac{1}{32\pi^2}\left[\frac{4\pi^2\lambda^2}{M^2}\right]^{\epsilon}\frac{\Gamma(1+\epsilon)}{M^2}
\end{equation}

\begin{equation}
i\int \frac{d^N p}{(2\pi)^N}\frac{p_{\mu}p_{\nu}}{(p^2-m^2)^2} = \frac{1}{32\pi^2}\left[\frac{4\pi^2\lambda^2}{M^2}\right]^{\epsilon}M^2\Gamma(\epsilon-1) g_{\mu\nu}
\end{equation}

\begin{equation}
i\int \frac{d^N p}{(2\pi)^N}\frac{p_{\mu}p_{\nu}}{(p^2-m^2)^3} = -\frac{1}{64\pi^2}\left[\frac{4\pi^2\lambda^2}{M^2}\right]^{\epsilon}\Gamma(\epsilon) g_{\mu\nu}
\end{equation}

\begin{equation}
i\int \frac{d^N p}{(2\pi)^N}\frac{p_{\mu}p_{\nu}p_{\rho}p_{\sigma}}{(p^2-m^2)^3} = \frac{1}{128\pi^2}\left[\frac{4\pi^2\lambda^2}{M^2}\right]^{\epsilon}M^2\Gamma(\epsilon-1) (g_{\mu\nu}g_{\rho\sigma}+g_{\mu\sigma}g_{\nu\rho}+g_{\mu\rho}g_{\nu\sigma})
\end{equation}

\end{document}